\documentclass[twocolumn,superscriptaddress,pre,letterpaper]{revtex4}

\usepackage[colorlinks=true, linkcolor=black, citecolor=blue, urlcolor=blue]{hyperref}
\usepackage{amsmath}
\usepackage{amsbsy}
\usepackage{fancybox}
\usepackage{mathtools}
\usepackage{bbm}
\usepackage{dcolumn}
\usepackage{enumitem}
\usepackage{graphicx}
\usepackage{float}
\usepackage{xcolor}
\usepackage{tikz}
\usepackage{mdframed}
\usetikzlibrary{arrows.meta}
\tikzset{state/.style={
    circle,
    draw,
    inner sep=0,
    minimum size=11mm,
    text width=11mm,
    align=center
}}
\tikzset{state2/.style={
    circle,
    draw,
    inner sep=0,
    minimum size=8mm,
    text width=8mm,
    align=center
}}

\newcommand\bm{\boldsymbol}

\newcommand{\e}{\mathrm{e}}
\newcommand{\dx}{\mathrm{d}x}
\newcommand{\dy}{\mathrm{d}y}
\newcommand{\du}{\mathrm{d}u}

\begin{document}
\title{Threshold and 
quasi-stationary distribution for the SIS model on networks}

\author{George Cantwell}
\email{gtc31@cam.ac.uk}
\affiliation{Department of Engineering, University of Cambridge, CB2 1PZ, United Kingdom}
\author{Cristopher Moore}
\email{moore@santafe.edu}
\affiliation{Santa Fe Institute, 1399 Hyde Park Road, Santa Fe, New Mexico 87501, USA}

\begin{abstract}
We study the Susceptible-Infectious-Susceptible (SIS) model on arbitrary networks.
The well-established \emph{pair approximation} treats neighboring pairs of nodes exactly while making a mean field approximation for the rest of the network.
We improve the method 
by expanding the state space dynamically, giving nodes a memory of when they last became susceptible. 
The resulting approximation is simple to implement and appears to be highly accurate, both in locating the epidemic threshold and in computing the quasistationary fraction of infected individuals above the threshold, for both finite graphs and infinite random graphs.
\end{abstract}

\maketitle

\section{Introduction}

The spread of an infectious disease can be modeled as a stochastic process unfolding on a network.
This perspective has significantly advanced our understanding of the mathematics of epidemics (e.g., \cite{kiss_mathematics_2017, pastor-satorras_epidemic_2015, keeling_networks_2005, grassberger_critical_1983, newman_spread_2002, moore_epidemics_2000,  Diekmann_1998}). 

Some diseases, such as measles or polio, infect individuals at most once (with rare exceptions).
Others, such as RSV or norovirus, can routinely reinfect the same individual.
The mathematical implications of this recurrent reinfection are quite subtle, since it creates ongoing epidemic dynamics on a network rather than sweeping through it just once.

The simplest epidemic models with reinfection are SIS models \cite{kiss_mathematics_2017, pastor-satorras_epidemic_2015, pastor-satorras_epidemic_2001, harris_contact_1974, weiss_asymptotic_1971}.
In these models, each individual is either susceptible (S) or infectious (I).
Susceptible individuals become infectious due to interactions with other infectious individuals, and infectious individuals recover and immediately become susceptible again.
The key finding of SIS models is that, under the right circumstances, a disease can enter an endemic phase where it continually infects people at a rate that balances out recovery, maintaining a finite fraction of infected individuals in the population. This endemic phase is quasi-stationary; it persists for exponential time until the system enters its absorbing state where no individuals are infected.

Even though the SIS model is exceedingly simple it is surprisingly resistant to analytic treatment.
Ideally we would like to locate the epidemic transition and define the properties of the endemic state, including the quasi-stationary fraction of infected individuals. 
But accurate characterization of either quantity has eluded the community, and remarkably we lack a solution even for the one-dimensional lattice.
Nevertheless, numerous approximate approaches have been advanced, e.g. \cite{kiss_mathematics_2017, pastor-satorras_epidemic_2015}. 

A key line of attack for SIS mirrors the development of methods to study percolation in the physics literature. In particular, the discrete-time SIS model is equivalent to a directed percolation model on a graph of nodes in space-time, with directed edges from one time step to the next corresponding to the spread or persistence of infection \cite{harris_contact_1974}. 

In order to compute epidemic thresholds, we typically start with simple mean-field  approximations where nodes are assumed to be independent~\cite{1238052, chakrabarti_epidemic_2008,  4549746, van2011n}. These approximations are inaccurate because they fail to account for correlations between neighbors. 

To go beyond the mean-field approximation, we can consider the joint distribution of pairs of connected individuals, while assuming that these pairs interact with each other in a mean-field way. This is known as the \emph{pair approximation} \cite{keeling_1999, mata_pair_2013, shrestha2015message, cator_second-order_2012, gleeson_2011}.
Unfortunately, the pair approximation is still noticeably inaccurate for SIS models, even when the network is locally tree-like and the corresponding approximation would be asymptotically exact for percolation \cite{karrer2010message, karrer_percolation_2014}.

In principle, one can extend the pair approximation by considering larger clusters of connected nodes, moving to connected triples, quadruples, and so on, but in practice this is rarely done. First, the computational effort of this approach grows exponentially with the size of the cluster. Second, while any network can easily be decomposed into a set of connected pairs, i.e., the edges, decomposition into larger motifs is non-trivial and non-unique (e.g., see~\cite{ferreira_critical_2013} for a triplet decomposition on regular graphs). 
Another approach along these lines is the star approximation \cite{gleeson_2013}, which yields a unique decomposition into overlapping star-shaped neighborhoods but at the expense of more complex equations.

We take a different approach in this paper.
Rather than extending the pair approximation in space, we extend it in time. 
The idea is reminiscent of the dynamic cavity method \cite{zhang_inference_2012,PhysRevE.91.012811,Dominguez_2020, Aurell_2023,PhysRevX.13.031021}, which has also been deployed to study recurrent disease models \cite{PhysRevE.105.024308}. These models work by keeping track of a node's entire history and then assuming that the histories of neighboring nodes are conditionally independent.
Tracking the  history of a node, however, requires computational effort that grows exponentially with the number of time steps or transitions. Thus this approach rapidly becomes intractable just as spatial cluster methods do. 

In contrast, we use a memoryless approach, but we augment the susceptible state to a chain of $K$ states that make transitions from one to the next according to a Poisson process. This provides an approximate ``clock'' that estimates how long a node has been susceptible, i.e., how long ago it was last infected.
These additional states let us model the correlation between this (approximate) time and the states of its neighbors. If a node was infected very recently then its neighbors are more likely to be infected, either because they infected it or vice versa. Conversely, if a node has been susceptible for a long time, its neighbors are less likely to be infected. 

We capture these correlations by assuming that the joint distribution of each neighboring pair of nodes is the stationary distribution of the corresponding Markov process. We then, analogous to the pair approximation, assume that pairs of nodes are conditionally independent of other pairs, taking the marginals of their shared endpoints into account.

By varying the number of states $K$, this approach yields a sequence of approximations for the SIS model that are simple to implement on arbitrary networks. 
While we remain agnostic on whether
this sequence approaches the exact behavior of the SIS model as $K \to \infty$, it converges quickly and is significantly more  accurate than standard methods.
Even using a chain of $K=8$ susceptible states gives results that agree very closely with experiment on a variety of networks.

\section{Approximating the SIS model}

We consider the continuous-time SIS model, also known as the contact process~\cite{harris_contact_1974}.
At any given time, each node is either susceptible or infectious. 
Each infectious node infects each of its susceptible neighbors at a constant rate, which we denote $\beta$.
Thus a susceptible node with $k$ infectious neighbors becomes infected at rate $\beta k$.
Once infectious, nodes recover at a constant rate and become susceptible once again.
Without loss of generality, we set the rate of recovery to $1$.

We will use the indicator variables $S_i$ and $I_i$ to denote the state of node~$i$ as susceptible or infectious respectively.
We can represent the model with the following diagram:
\begin{equation}
  \begin{tikzpicture}
    \node[state] (Si) at (0,0) {$S_i$};
    \node[state] (Sj) at (3.5,0) {$I_i$};
	  \draw[-{Stealth[length=7pt, width=5pt]}] (Si) to[bend left=10] node[midway, above] {\(\beta \sum_{j} A_{ij} I_j \)} (Sj);
    \draw[-{Stealth[length=7pt, width=5pt]}] (Sj) to[bend left=10] node[midway, below] {\( 1 \)} (Si);
  \end{tikzpicture} \nonumber 
\end{equation}
where 
$A$ is the adjacency matrix so that the sum $\sum_{j}A_{ij} I_j$ is the number of infectious neighbors of node~$i$.

\subsection{The individual approximation}

While the system as a whole is Markovian, any sub-system is not.
Nevertheless, in the style of a mean-field approximation, let us assume that each node follows its own Markov process, becoming infected at a rate proportional to its average number of infectious neighbors.
Diagrammatically, this becomes
\begin{equation}
  \begin{tikzpicture}
    \node[state] (Si) at (0,0) {$S_i$};
    \node[state] (Sj) at (3.5,0) {$I_i$};
	  \draw[-{Stealth[length=7pt, width=5pt]}] (Si) to[bend left=10] node[midway, above] {\(\beta \sum_{j} A_{ij} \langle I_j \rangle \)} (Sj);
    \draw[-{Stealth[length=7pt, width=5pt]}] (Sj) to[bend left=10] node[midway, below] {\( 1 \)} (Si);
  \end{tikzpicture}. \nonumber 
\end{equation}
The stationary distribution for node~$i$ is 
\begin{equation}
	P_{i}(I_i) = \langle I_i \rangle = \frac{\beta \sum_j A_{ij} \langle I_j \rangle }{1+\beta \sum_j A_{ij} \langle I_j \rangle }
    \, .
\end{equation}
or in matrix notation we can write
\begin{equation}
	\bm{\rho} = \frac{\beta A \bm{\rho} }{1+\beta A \bm{\rho} }
    \label{eq:mean-field}
\end{equation}
where $\rho_i = \langle I_i \rangle = P_{i}(I_i)$.
This simple set of self-consistent equations provides a qualitatively reasonable  approximation to the full model. Linearizing~\eqref{eq:mean-field} around $\bm{\rho}=\bm{0}$ leads to the first-order approximation for the epidemic threshold, namely
\begin{equation}
\beta_{\mathrm{c}} \simeq \beta_{\mathrm{c}}^{(0)} = \frac{1}{\lambda(A)}
\end{equation}
where $\lambda(A)$ is the largest eigenvalue of $A$ (e.g., \cite{1238052, chakrabarti_epidemic_2008,  4549746}).

A regular graph provides the simplest example.
If every node has $q+1$ neighbors, as in a tree where each node has $q$ children and a parent, then Eq.~\eqref{eq:mean-field} predicts
\begin{equation}
  \rho_i = \frac{\beta (q+1) - 1}{\beta (q+1)}
\end{equation}
when $\beta > \beta_{\mathrm{c}}^{(0)} = 1/(q+1)$, and $\rho_i = 0$ otherwise.
However, both Eq.~\eqref{eq:mean-field} and this estimate for $\beta_{\mathrm{c}}$ are inaccurate, as we shall presently see.

\subsection{The pair approximation}

To improve the approximation of the previous section, let us instead assume that each connected \emph{pair} of nodes is Markovian.
For each connected pair $(i,j)$, we will endeavor to find $P_{ij}(X_i, X_j)$, the joint probability that that node~$i$ is in state $X_i$ and $j$ in state $X_j$.
To do this, we will (incorrectly) assume that the small sub-system of $(i,j)$ is Markovian and that infections from outside the pair arrive at their average rate.
When node~$j$ is susceptible, it will be infected by nodes other than $i$ at rate
\begin{equation}
	\beta \sum_{k \neq i} A_{jk} P_{jk}(I_k \vert S_j) \, .
\end{equation}
Defining
\begin{equation}
\phi_{jk} = \beta P_{jk}(I_k \vert S_j)
\end{equation}
then the rate of infection from outside the pair can be written compactly using operator notation.
Namely, by defining
\begin{equation}
  B_{ij}(\boldsymbol{\phi}) = A_{ij} \sum_{k \neq i} A_{jk} \phi_{jk} 
  \, , 
\end{equation}
the rate of infection from outside the pair is $B_{ij}(\bm{\phi})$ for node $j$ and $B_{ji}(\bm{\phi})$ for node $i$.
The operator $B$ is the non-backtracking operator (e.g.,~\cite{spectral_redemption_2013})
and keeps track of how the infection spreads to the pair due to the rest of the network.
Diagrammatically, the picture we get is:
\begin{equation}
  \begin{tikzpicture}
    \node[state] (SS) at (0,0) {$S_i S_j$};
    \node[state] (SI) at (3.3,0) {$S_i I_j$};
    \node[state] (IS) at (0,-3) {$I_i S_j$};
    \node[state] (II) at (3.3,-3) {$I_i I_j$};

    \draw[-{Stealth[length=7pt, width=5pt]}] (SI) to[bend left=10] node[midway, below] {\( 1 \)} (SS);
    \draw[-{Stealth[length=7pt, width=5pt]}] (SS) to[bend left=10] node[midway, above] {\( B_{ij}(\bm{\phi}) \)} (SI);

    \draw[-{Stealth[length=7pt, width=5pt]}] (II) to[bend left=10] node[midway, left] {\( 1 \)} (SI);
    \draw[-{Stealth[length=7pt, width=5pt]}] (SI) to[bend left=10] node[midway, right] {\(\beta + B_{ji}(\bm{\phi}) \)} (II);

    \draw[-{Stealth[length=7pt, width=5pt]}] (IS) to[bend right=10] node[midway, right] {\( 1 \)} (SS);
    \draw[-{Stealth[length=7pt, width=5pt]}] (SS) to[bend right=10] node[midway, left] {\( \qquad B_{ji}(\bm{\phi}) \)} (IS);

    \draw[-{Stealth[length=7pt, width=5pt]}] (II) to[bend right=10] node[midway, above] {\( 1 \)} (IS);
    \draw[-{Stealth[length=7pt, width=5pt]}] (IS) to[bend right=10] node[midway, below] {\(\beta  + B_{ij}(\bm{\phi}) \)} (II);

  \end{tikzpicture}. \nonumber 
\end{equation}
If we knew $\bm{\phi}$, we could solve this simple system for its stationary distribution following elementary methods \cite{Norris_1997}.
To do this one solves for the left-eigenvector with zero eigenvalue of the transition matrix
\begin{equation}
\begin{aligned}
	Q(B_{ij}(\bm{\phi}), B_{ji}&(\bm{\phi})) = \\ 
	&\left(
\begin{array}{cccc}
\cdot & 1 & 1 & 0 \\
\beta +  B_{ij}(\bm{\phi}) & \cdot & 0 & 1 \\
\beta +  B_{ji}(\bm{\phi}) & 0 & \cdot & 1 \\
0 & B_{ji}(\bm{\phi}) &  B_{ij}(\bm{\phi}) & \cdot
\end{array}
\right)
\end{aligned}
\end{equation}
where the diagonals are set so that the rows sum to zero.
This provides an expression for the distribution $P_{ij}$ given knowledge of $\bm{\phi}$. Namely, $P_{ij}$ is the vector that solves
\begin{equation}
	P_{ij} Q(B_{ij}(\bm{\phi}), B_{ji}(\bm{\phi})) = \bm{0}.
	\label{eq:SIS1_matrix}
\end{equation}

Of course the whole point of this calculation is that we do not know $\bm{\phi}$ in advance.  Regardless, from its definition we write
\begin{equation}
	\phi_{ij} = \beta P_{ij}(I_j|S_i) = \frac{\beta P_{ij}(S_i, I_j)}{P_{ij}(S_i, I_j) + P_{ij}(S_i, S_j)}.
  \label{eq:phi}
\end{equation}
With this, we have a closed system of equations. Eq.~\eqref{eq:SIS1_matrix} gives us the required probabilities in terms of $\bm{\phi}$ and Eq.~\eqref{eq:phi} gives us $\bm{\phi}$ in terms of the probabilities.
By explicitly inverting the matrix in Eq.~\eqref{eq:SIS1_matrix} and inserting the resulting distribution into the definition of Eq.~\eqref{eq:phi}, we find 
\begin{equation}
	\phi_{ij} = \Psi_{\beta}\big(B_{ij}(\bm{\phi}), B_{ji}(\bm{\phi}) \big)
	\label{eq:pair_approx}
\end{equation}
with
\begin{equation}
	\Psi_{\beta}(x, y) = \beta \, \frac{2x + x y + x^2 + \beta x + \beta y}{(2+x+y)(1+x+\beta)}.
\end{equation}
On any network of interest this set of equations, Eq.~\eqref{eq:pair_approx}, can be iterated until convergence to find $\bm{\phi}$.

Finally, when we want to approximate the behavior of a single node~$i$, we can assume it receives infections from each neighbor~$j$ at rate $\phi_{ij}$, and thus
\begin{equation}
	P_i(I_i) = \frac{\sum_j A_{ij} \phi_{ij}}{1 + \sum_j A_{ij} \phi_{ij}}
    \, .
\end{equation}

While we have derived this via a conceptually different route, this approximation is equivalent to the standard pair approximation~\cite{mata_pair_2013, pastor-satorras_epidemic_2015}.
The main difference is that in the conventional set-up for the pair approximation, one integrates a large system of nonlinear equations, whereas by solving Eq.~\eqref{eq:pair_approx} we directly jump to the steady-state solution without needing to numerically integrate.

As seen above, the single-node mean-field approximation predicts that the endemic transition occurs at $\beta = 1 / \lambda(A)$.
The pair approximation improves on this estimate.
Linearizing Eq.~\eqref{eq:pair_approx} at $\phi_{ij}=0 + \varepsilon_{ij}$ we get
\begin{equation}
\varepsilon_{ij} = \frac{\beta(\beta+2)}{2(\beta+1)} B_{ij}(\boldsymbol{\varepsilon}) + \frac{\beta^2}{2(\beta+1)} B_{ji}(\boldsymbol{\varepsilon})
\label{eq:linear_Psi}
\end{equation}
which changes stability at a critical value $\beta_{\mathrm{c}}^{(1)}$, the solution to
\begin{equation}
\beta_{\mathrm{c}}^{(1)} = \frac{1}{\lambda \big(A-\beta_{\mathrm{c}}^{(1)}L/2 - I\big)}
\end{equation}
where $L = D - A$ is the graph Laplacian (see Appendix \ref{app:evalue}).
This fixed point equation can be easily solved numerically by iteration.

For a $(q+1)$-regular random graph, Eq.~\eqref{eq:pair_approx}  becomes
\begin{equation}
	\phi = \Psi_{\beta}( q \phi, q \phi ) = \frac{q \beta \phi}{1 + q \phi} 
    \, .
\end{equation}
This yields $\phi = \beta - 1/q$, 
so this approximation predicts that the endemic threshold is  
\begin{equation}
\label{eq:beta-1}
\beta_{\mathrm{c}}^{(1)} = 1/q \, .
\end{equation}
While this is not a bad prediction, it disagrees with simulations by an appreciable margin, particularly for small $q$. Below we give a correction derived from our approach.

\subsection{Improved approximation for dynamical correlations}

Having reviewed the mean-field and pair approximations, we turn now to our augmentation of the state space.
We consider a family of variations of the SIS model, which we call the SIS\textsuperscript{$K$} model, with $K \geq 1$.
The SIS\textsuperscript{$1$} model is the SIS model.
The SIS\textsuperscript{$2$} model is very similar to the SIS model, except that there are $2$ different $S$ states, $S^{(1)}$ and $S^{(2)}$.
The transitions are
\begin{equation}
  \begin{tikzpicture}
    \node[state2] (S1) at (0,0) {$S^{(1)}_i$};
    \node[state2] (S2) at (0,-2.) {$S^{(2)}_i$};
    \node[state2, minimum size=9mm] (I) at (1.732,-1.) {$I_i$};

    \draw[-{Stealth[length=7pt, width=5pt]}] (S2) to[bend left=0] node[midway, left] {\( \gamma \)} (S1);
    \draw[-{Stealth[length=7pt, width=5pt]}] (S1) to[bend left=0] node[pos=0.1, midway, above, xshift=-4] {\( \qquad\qquad \beta \sum_j A_{ij} I_j \)} (I);
    \draw[-{Stealth[length=7pt, width=5pt]}] (S2) to[bend right=10] node[midway, below] {\( \qquad\qquad \beta \sum_j A_{ij} I_j \)} (I);
    \draw[-{Stealth[length=7pt, width=5pt]}] (I) to[bend right=10] node[pos=0.5, midway, above] {\( 1 \)} (S2);
  \end{tikzpicture}. \nonumber 
\end{equation}
Just as for the standard SIS model, the infectious state $I$ recovers at unit rate.
Both $S^{(1)}$ and $S^{(2)}$ are infected at exactly the same rate as the standard SIS model.
State $I$ recovers to $S^{(2)}$, and then $S^{(2)}$ decays to $S^{(1)}$ at some rate $\gamma$. 

Clearly nodes in state $S^{(1)}$ have been susceptible for longer, on average, than nodes in state $S^{(2)}$. However, this variant model is ultimately the same as the conventional SIS model: if we collapse or ``lump'' the states $S^{(1)}$ and $S^{(2)}$ together in a single $S$ state, we recover the original SIS dynamics exactly. 
 
The variants SIS\textsuperscript{$K$} with $K > 2$ are defined in the same way.
We have $K$ different $S$ states, $S^{(1)}, \dots, S^{(K)}$.
Each is infected at exactly the same rate; $I$ always recovers to $S^{(K)}$; and each $S^{(x+1)}$ decays to $S^{(x)}$ at rate $\gamma$. $S^{(1)}$ does not decay further, so its only transition is to become infectious.
The picture is now
\begin{equation}
  \begin{tikzpicture}
    \node[state2] (S1) at (0,0) {$S^{(1)}_i$};
    \node[state2] (S2) at (-2,0) {$S^{(2)}_i$};
    \node (S3) at (-3.6,0) {$\dots$};
    \node[state2] (SK) at (-5.2,0) {$S^{(K)}_i$};
    \node[state2] (I) at (-7.2,0) {$I_i$};

    \draw[-{Stealth[length=7pt, width=5pt]}] (S2) to[bend left=0] node[midway, above] {\( \gamma \)} (S1);
    \draw[-{Stealth[length=7pt, width=5pt]}] (S3) to[bend left=0] node[midway, above] {\( \gamma \)} (S2);
    \draw[-{Stealth[length=7pt, width=5pt]}] (SK) to[bend left=0] node[midway, above] {\( \gamma \)} (S3);
    \draw[-{Stealth[length=7pt, width=5pt]}] (I) to[bend left=0] node[midway, above] {\( 1 \)} (SK);
    \draw[-{Stealth[length=7pt, width=5pt]}] (S1) to[bend left=55] node[pos=0.15, below, yshift=1pt, xshift=18pt] {\(\beta \sum_j A_{ij} I_j \)} (I);
    \draw[-{Stealth[length=7pt, width=5pt]}] (S2) to[bend left=45] node[pos=0.18, below, yshift=1pt, xshift=18pt] {\(\beta \sum_j A_{ij} I_j \)} (I);
    \draw[-{Stealth[length=7pt, width=5pt]}] (SK) to[bend left=30] node[pos=0.15, below, yshift=-1pt, xshift=10pt] {\( \beta \sum_j A_{ij} I_j \)} (I);
  \end{tikzpicture} \nonumber .
\end{equation}
Again, the key point of this construction is that for all $K$, SIS\textsuperscript{$K$} is exactly equivalent to SIS if we lump the $S^{(1)},\dots,S^{(K)}$ together into a single $S$ state \footnote{This construction is essentially opposite to the so-called linear chain trick \cite{MacDonald78}. Because real-world recovery times are not well modeled by an exponential distribution, more accurate epidemic models can be constructed by incorporating extra states, so that the total time it takes to recover is, e.g., Gamma-distributed. Thus the linear chain trick deliberately changes the behavior of the model, whereas our additional states are constructed so that they make no change to the model whatsoever.}. 

Because  SIS\textsuperscript{$K$} has precisely the same behavior as the standard SIS model, the extra $S$ states may seem like an unnecessary complication.
A simple appeal to Occam's razor would have us do away with them.
However, they turn out to be a useful conceptual tool.
Their true utility becomes apparent once we start to approximate the system, rather than performing exact simulations.

While an exact simulation of SIS\textsuperscript{$K$} and the standard SIS model are equivalent, the approximations to them are not.
Let us follow the same logic as in the pair approximation above.
If nodes~$i$ and $j$ are connected then we will solve the equilibrium dynamics of the pair exactly, while assuming they are infected by their other neighbors at rate that depends on their state.
When node~$j$ is in state $S^{(x)}$, it is infected by its other neighbors at rate
\begin{equation}
	\beta \sum_{k \neq i} A_{jk} P_{jk}(I_k \vert S_j^{(x)}) \, .
\end{equation}
Defining 
\begin{equation}
\phi_{jk}^{x} = \beta P_{jk}(I_k \vert S_j^{(x)}) \, ,
\end{equation}
we have the rate of infection for node $j$ from neighbors other than $i$ to be $B_{ij}(\bm{\phi}^x)$
for each $x=1,\dots,K$.

Each node in the pair $(i,j)$ can be in any one of $K+1$ states and so the system consisting of the pair can be in any of \mbox{$(K+1)^2$} states.
Let $Q(\boldsymbol{a},\boldsymbol{b})$ denote the transition matrix for a system of two nodes with infections from outside the pair occurring at rate $a_x$ (resp.\ $b_x$) when the first (resp.\ second) node is in state $S^{(x)}$.
The elements of the transition matrix are, for recovery events,
\begin{equation}
\begin{aligned}
	  \left[Q(\boldsymbol{a},\boldsymbol{b})\right]_{II, IK} 
	&= \left[Q(\boldsymbol{a},\boldsymbol{b})\right]_{II, KI} \\
	&= \left[Q(\boldsymbol{a},\boldsymbol{b})\right]_{Ix, Kx} \\
	&= \left[Q(\boldsymbol{a},\boldsymbol{b})\right]_{xI, xK} 
	= 1 \, ,
\end{aligned}
\end{equation}
for the decay events
\begin{equation}
\begin{aligned}
	\left[Q(\boldsymbol{a},\boldsymbol{b})\right]_{Ix, I(x-1)} 
	&= \left[Q(\boldsymbol{a},\boldsymbol{b})\right]_{xI, (x-1)I} \\
	&= \left[Q(\boldsymbol{a},\boldsymbol{b})\right]_{xy, (x-1)y} \\
	&= \left[Q(\boldsymbol{a},\boldsymbol{b})\right]_{xy, x(y-1)} = \gamma \, ,
\end{aligned}
\end{equation}
and for infection events 
\begin{align}
	\left[Q(\boldsymbol{a},\boldsymbol{b})\right]_{Iy, II} &= \beta + b_y \\
	\left[Q(\boldsymbol{a},\boldsymbol{b})\right]_{xI, II} &= \beta + a_x \\
	\left[Q(\boldsymbol{a},\boldsymbol{b})\right]_{xy, xI} &= b_y \\
	\left[Q(\boldsymbol{a},\boldsymbol{b})\right]_{xy, Iy} &= a_x \, .
\end{align}
As usual, the diagonals of the transition matrix $Q$ are set so that all rows sum to zero.

Assuming we knew the rates $\phi_{ij}^x$,
the stationary distribution for the system can be found by solving a linear equation,
namely the distribution $P_{ij}$ solves
\begin{equation}
	P_{ij}  \, Q\big(B_{ij}(\bm{\phi}), B_{ji}(\bm{\phi}) \big) = \bm{0} \,.
	\label{eq:transition_eq}
\end{equation}
On the other hand, if we knew the stationary distribution $P_{ij}$ we could easily compute
\begin{equation}
	\phi_{ji}^{x} = \frac{\beta \, P_{ij}(I_i, S_j^{(x)}) }{ P_{ij}(I_i, S_j^{(x)}) + \sum_{y=1}^{K} P_{ij}(S_i^{(y)}, S_j^{(x)})  }
    \, . \label{eq:message_eq}
\end{equation}
Hence, we are again left with a set of self-consistent equations that we can solve by iteration.
We alternately find $P_{ij}$ for all connected pairs $(i,j)$ by solving the linear equation in Eq.~\eqref{eq:transition_eq} and then we update our estimates of $\phi_{ji}^x$ using Eq.~\eqref{eq:message_eq}.

The extra $S$ states effectively track how recently a node was infectious.
At very short times after moving from infectious back to susceptible, a node will likely be in state $S^{(K)}$.
At longer times, if it has not been reinfected, it will eventually decay back to $S^{(1)}$.
Which $S$ state a node is in thus acts like a noisy clock that ``ticks'' from $S^{(x+1)}$ to $S^{(x)}$ at rate $\gamma$.
More precisely, the time taken to decay from $S^{(x+\gamma t)}$ to $S^{(x)}$ is a (Gamma distributed) random variable with mean $t$, while the variance is $t / \gamma$.
Hence the ``clock'' measures time at a resolution inversely proportional to $\gamma$ and the maximum time that these $K$ states can record in this sense is $K/\gamma$. 

We can tune the decay rate $\gamma$ as we wish, in order for the state distribution to be as informative as possible about how long a node has been susceptible.
In order for both the temporal resolution and the maximum time to improve as $K$ grows, we propose setting $\gamma$ to
\begin{equation}
	\gamma = \beta q \sqrt{K - 1}
    \, ,
	\label{eq:gamma}
\end{equation}
where $q$ is one less than the mean degree, i.e., $q = \sum_{ij}A_{ij}/n - 1$, and we use $\sqrt{K-1}$ because there are $K-1$ transitions from $S^{(K)}$ to $S^{(1)}$. 
This choice clearly meets our temporal scaling requirements: $1/\gamma$ tends to zero as $K$ gets large and additionally $K/\gamma$ tends to infinity, and thus we obtain a more fine-grained resolution over a wider range of times as $K$ increases. Of course, using $\sqrt{K}$ would work as well.

Additionally, note that the number of infections arriving at a node per unit time scales linearly with $\beta q$, whereas the number of decay events at a node (where $S^{(x+1)}$ decays to $S^{(x)}$) scales linearly with $\gamma$.
For these to scale equivalently, and hence for the timescales of infections and decays to match, we require that $\gamma$ is proportional to $\beta q$.
Equation~\eqref{eq:gamma} matches our desiderata, although of course it is not the only such choice.

\subsubsection{Dynamic correlation}
The standard pair approximation allows us to approximate spatial correlations, i.e., $\langle I_i I_j \rangle = P_{ij}(I_i, I_j)$.
Our SIS\textsuperscript{$K$} pair approximation also allows us to compute this and if it approximates $P_{ij}$ more accurately, then it should approximate spatial correlations more accurately.
However, SIS\textsuperscript{$K$} additionally allows us to improve on our approximations for dynamic (a.k.a.\ temporal) correlations.

The SIS\textsuperscript{$K$} approximation is built by considering each connected pair of nodes to be a small Markovian system.
As a result, we can ask questions such as: what is the probability that a node will remain susceptible for a given length of time?
To answer such questions, we first solve the self-consistent equations that define our approximation, i.e., we solve for
\begin{equation}
	\phi_{ij}^{x} = \beta P_{ij}(I_j \vert S_i^{(x)})
    \, .
\end{equation}
Once we know $\phi_{ij}^{x}$ we can treat node~$i$ as an isolated Markovian system which, when in state $S^{(x)}$, is infected at rate
\begin{equation}
	\lambda_i^{(x)} = \sum_{j} A_{ij} \phi_{ij}^{x}
    \, .
\end{equation}
This yields a $(K+1)$-state Markov process described by the following diagram:
\begin{equation}
  \begin{tikzpicture}
    \node[state2] (S1) at (0,0) {$S^{(1)}_i$};
    \node[state2] (S2) at (-2,0) {$S^{(2)}_i$};
    \node (S3) at (-3.6,0) {$\dots$};
    \node[state2] (SK) at (-5.2,0) {$S^{(K)}_i$};
    \node[state2] (I) at (-7.2,0) {$I_i$};

    \draw[-{Stealth[length=7pt, width=5pt]}] (S2) to[bend left=0] node[midway, above] {\( \gamma \)} (S1);
    \draw[-{Stealth[length=7pt, width=5pt]}] (S3) to[bend left=0] node[midway, above] {\( \gamma \)} (S2);
    \draw[-{Stealth[length=7pt, width=5pt]}] (SK) to[bend left=0] node[midway, above] {\( \gamma \)} (S3);
    \draw[-{Stealth[length=7pt, width=5pt]}] (I) to[bend left=0] node[midway, above] {\( 1 \)} (SK);
    \draw[-{Stealth[length=7pt, width=5pt]}] (S1) to[bend left=55] node[pos=0.15, below, yshift=1pt] {\(\lambda_{i}^{(1)} \)} (I);
    \draw[-{Stealth[length=7pt, width=5pt]}] (S2) to[bend left=45] node[pos=0.15, below, yshift=1pt] {\(\lambda_{i}^{(2)} \)} (I);
    \draw[-{Stealth[length=7pt, width=5pt]}] (SK) to[bend left=30] node[pos=0.15, below, yshift=1pt] {\(\lambda_{i}^{(K)} \)} (I);
  \end{tikzpicture} .
  \nonumber
\end{equation}

It is straightforward to solve for the stationary distribution of this process. We can then compute quantities such as the distribution of times between infections.
At the moment when node~$i$ ceases to be infectious, it will be in state $S^{(K)}$.
At some later time $\Delta_I$ it will be infectious again as long as we are above the endemic threshold.
This inter-infection time $\Delta_I$ will be a random variable, and we can compute its distribution as follows. 
Let $R$ be the transition matrix $R$ of this one-node Markov process with $I$ as an absorbing state: i.e., $[R(\boldsymbol{\lambda})]_{x,x-1} = \gamma$,
$[R(\bm{\lambda})]_{x,I} = \lambda_x$, and diagonals are set so that row sums are zero. Then the matrix exponential $\e^{t R(\boldsymbol{\lambda}_i)}$ 
describes the distribution after being susceptible for time $t$, and its entry
\begin{equation}
[\e^{t R(\boldsymbol{\lambda}_i)}]_{K,I} = P(\Delta_I \leq t)
\end{equation}
gives the probability of being infected after time $t$ from the initial state $S^{(K)}$.

\subsubsection{Random regular graphs}

For the SIS\textsuperscript{$2$} model on a random $(q+1)$-regular graph, we again have that  every node and every edge is equivalent, and so $\phi_{ij}^{x} = \phi_{x}$ for all $ij$ and $B_{ij}(\bm{\phi}^{x}) = q \phi_x$.
Inserting these into to the $9 \times 9$ transition matrix, after considerable algebra, we arrive at the the self-consistent equation
\begin{widetext}
\begin{equation}
\begin{pmatrix}
\phi_1 \\
\phi_2	
\end{pmatrix} = 
\begin{pmatrix}
\frac{\beta  q \phi _1 \left(\beta +\phi _2\right) \left(\beta  q+q \left(\phi _1+\phi
   _2\right)+1\right)}{q \phi _1^2 \left(\beta  q+q \phi _2+1\right)+\phi _1 \left(\beta
   +q \phi _2 \left(2 \beta  q+q \phi _2+2\right)+\beta  q (\beta  q+3)+1\right)+\beta 
   \left(\beta +\beta  q+q \phi _2+1\right)} \\
\frac{\beta  q \left(\beta +\phi _2\right) \left(\beta  \left(\beta +\phi _2\right)+\phi
   _1 \left(\beta +\beta  q+q \phi _2+1\right)+q \phi _1^2\right)}{q \phi _1^2
   \left(\beta  q+q \phi _2+1\right)+\phi _1 \left(\beta +q \phi _2 \left(\beta +2 \beta 
   q+q \phi _2+2\right)+\beta  q (\beta +\beta  q+3)+1\right)+\beta  \left(\phi _2
   \left(2 \beta  q+q \phi _2+q+1\right)+\beta  (\beta  q+q+2)+1\right)}
\end{pmatrix}
\end{equation}
\end{widetext}
This equation has a trivial solution at $\phi_1=0$.  
The trivial point changes stability at the critical value of
\begin{equation}
	\beta_{\mathrm{c}}^{(2)} = \frac{1 -2 q^3+4 q+(2 q+1) \sqrt{4 q (q+1)^3+1}}{2 q^2 (q+2)^2}
	\label{eq:beta_c^2}
\end{equation}
and to leading order in $1/q$ we find
\begin{equation}
\label{eq:beta-2}
\beta_{\mathrm{c}}^{(2)} \approx \frac{1}{q} + \frac{1}{4 q^3} + \mathcal{O}\Big( \frac{1}{q^4} \Big) \, .
\end{equation}

This prediction for $\beta_c$ improves on the standard result $\beta_\mathrm{c}^{(1)} = 1/q$ from the pair approximation. Interestingly it also predicts that the first correction to $\beta_c$ in a high-degree expansion is $O(1/q^3)$.
We conjecture that this prediction is correct, although we do not believe that $1/4$ is the correct coefficient (see Sec~\ref{sec:SIS_inf} for further discussion).
Regardless, Eq.~\eqref{eq:beta-2} gives a noticeable improvement when compared to simulation. 
For example, for the $3$-regular random graph where $q=2$ we numerically find $\beta_\mathrm{c} \approx 0.545$, 
whereas the theoretical predictions are
$\beta_\mathrm{c}^{(0)} = 1/3$, $\beta_\mathrm{c}^{(1)} = 1/2$, and $\beta_\mathrm{c}^{(2)} = 0.521$. 

Having motivated the 
SIS\textsuperscript{$K$} 
model and derived its pair approximation,
in the next section we present numerical experiments to demonstrate its accuracy.

\section{Numerical experiments}

To approximate the SIS model on an arbitrary network, we fix a number $K \geq 1$ of  $S$ states and then solve the set of self-consistent equations \eqref{eq:transition_eq} and \eqref{eq:message_eq} for $\phi^{x}_{jk} = \beta P(I_k \vert S_j^{(x)})$. 
Note that this procedure can be carried out on a finite network, or analyzed asymptotically on large random graphs \`a la the cavity method.

\begin{figure}
\centering
\includegraphics{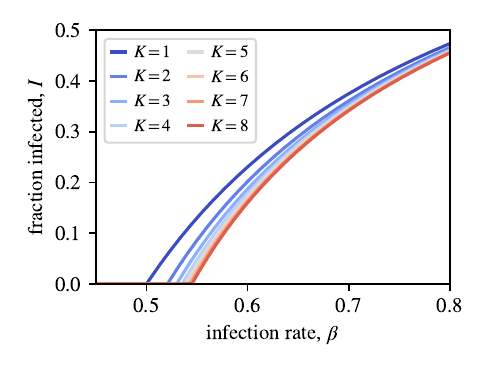}
\caption{\label{fig:convergence_K} Approximations for the SIS model on $3$-regular random graphs with increasing $K$.
We show the predicted steady-state fraction of infectious nodes as a function the spreading rate, $\beta$.
As $K$ increases we see a diminishing change to the predictions at $K$ against $K-1$.
Note that throughout this paper we set the recovery rate to $1$ without loss of generality.
}
\end{figure}

\subsection{Random regular graphs}
Increasing $K$ increases the size of the system of linear equations we must solve in Eq.~\eqref{eq:transition_eq} and also changes the predictions.
To explore this, we carried out our approximation for $K=1,2,\dots,8$ on $3$-regular random graphs.
Results are shown in Fig.~\ref{fig:convergence_K} and show a diminishing effect of increasing $K$. 
This suggests the approximation is well-behaved with $K$ and converges reasonably quickly.

\begin{figure}
\centering
\includegraphics{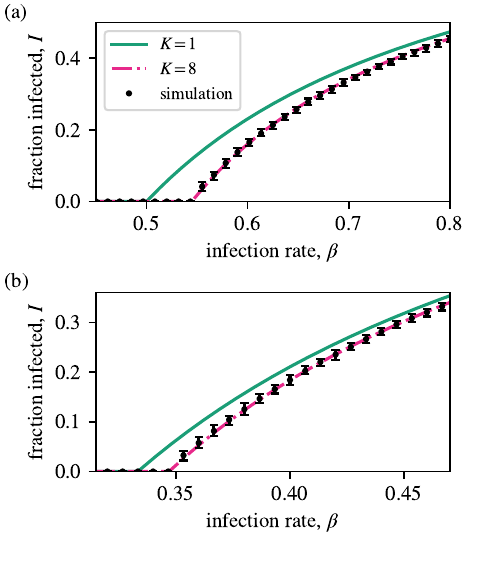}
\caption{\label{fig:frac_infected_sims} Fraction of infected nodes at different rates of infection on (a) $3$-regular random graphs and (b) $4$-regular random graphs. We compare simulations on a $50,00$ node network to our approximations with $K=1$ and $K=8$. We see a compelling improvement in the predictions.}
\end{figure}

Next, and arguably more importantly, we compare the approximation to Monte Carlo simulations.
In Fig.~\ref{fig:frac_infected_sims} we simulate the SIS process on large $3$- and $4$-regular random graphs at varying $\beta$.
We run the simulations for $10,000$ units of time on graphs of size $N=50,000$; this should be sufficiently long and sufficiently large to minimize transient and finite-size effects. 
We find that the $K=8$ approximation is practically indistinguishable from simulation results, with respect to both the location of the transition and the proportion infected in the endemic state above the transition.

\begin{figure}
\centering
\includegraphics{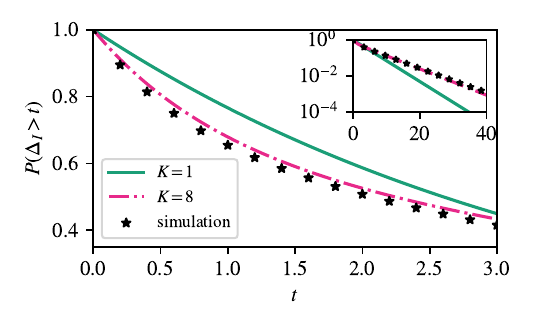}
\caption{\label{fig:survival_fn} Estimates of the survival function for a $4$-regular random graph.
Let $\Delta_I$ be the time between infections in the endemic state.
We show the probability that $\Delta_I > t$, i.e., that the node survives in the susceptible state for longer that $t$ between infections.
Simulations are on a graph with $50,000$ nodes and $\beta=0.4$. The inset shows the same data but on a logarithmic scale.
Setting $K=1$, i.e., using the standard pair approximation, yields a simple exponential, while the $K=8$ approximation closely matches experiment.}
\end{figure}

As discussed above, by increasing $K$ an effective memory is built into the system and our answers to temporal questions should also improve in accuracy.
We see exactly this in Fig.~\ref{fig:survival_fn}.
We simulated the SIS model on a large $4$-regular random graph at $\beta=0.4$ (which is above the threshold) and we measured the time between infections.
From these measurements, we can estimate the probability that a node will survive in the $S$ state for more than $t$ time units.
Note that by the standard approximation, the survival time distribution is a simple exponential. Thus our goal is to correctly model how correlations create a mix of exponentials. In particular, susceptible nodes that were recently infected are more likely to become reinfected quickly, so the risk of reinfection decreases as $t$ increases.

We show the survival probability on both linear and logarithmic scales in Fig.~\ref{fig:survival_fn}, and compare simulations to the $K=1$ and $K=8$ approximations. While setting $K=1$ yields a simple exponential, the $K=8$ approximation is quite accurate, closely matching both the curvature and tail of the survival time distribution.

\subsection{Real-world examples}

\begin{figure*}
\centering
\includegraphics[width=1.75\columnwidth]{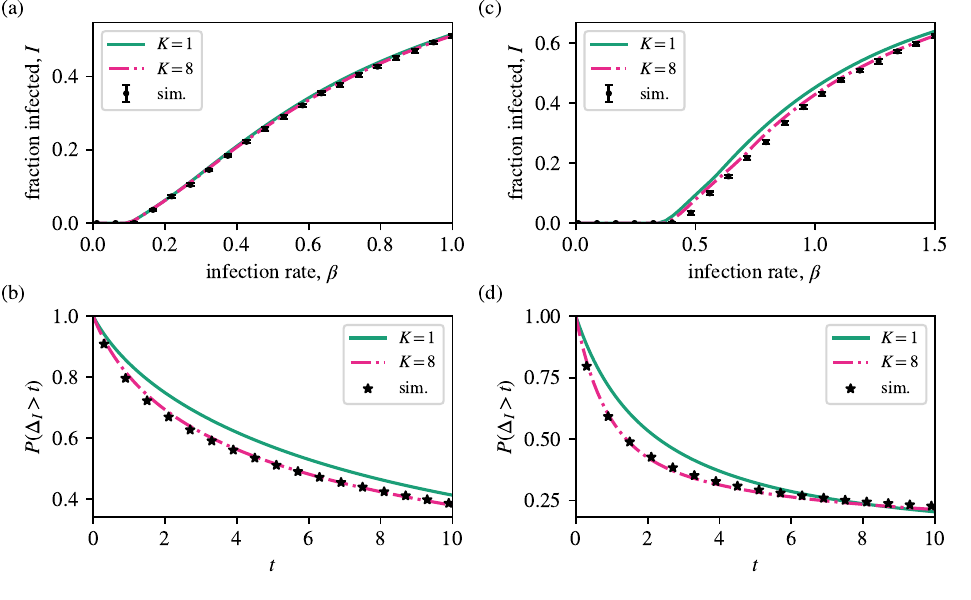}
\caption{\label{fig:real_world_experiments} Simulations on real-world  networks. 
Panels (a) and (b) are the results of simulations on a sexual contact network of Ref.~\cite{morris_hiv_2011}.
Panels (c) and (d) are the results of simulations on a network of roads, see Ref. \cite{subelj_robust_2011}.
In (a) and (c) we show the fraction infected after simulating for $10,000$ units of time at varying values of $\beta$. 
In panels (b) and (d) we show the survival function at $\beta = 0.4$ and $\beta = 0.6$ respectively.
}
\end{figure*}

The uniform nature of random regular graphs makes our approximation very simple, since every edge is equivalent. 
However, unlike some other higher-order approximations, our approximations can be directly applied to arbitrary graphs.
For any value of $K$, our approximation provides a system of linear equations for each edge in the network, and the set of all these equations can be solved by a simple fixed-point iteration scheme.

To explore how our approximations behave on more realistic structures, we compute them for two real-world network data sets.
The first of these is a sexual contact network \cite{morris_hiv_2011}, while the second is a network of European roads \cite{subelj_robust_2011}.
In Fig.~\ref{fig:real_world_experiments} we show the results of our approximation and again these are compared with direct simulation of the SIS process. 
As in random regular graphs, our approximation quickly improves and converges as $K$ increases while $K=8$ closely agrees with our simulation.

\section{The $K \to \infty$ limit}
\label{sec:SIS_inf}

In our experiments so far, we found that the accuracy of our approximation increases quickly as $K$ increases.
On the other hand, increasing $K$ also increases the complexity of our method, since in each iteration we need to solve a system of $L=(K+1)^2$ linear equations to compute the stationary distribution of each pair, and doing this with standard linear algebra requires $O(L^3)=O(K^6)$ operations. While this growth in complexity is only polynomial, and we might be able to decrease it by exploiting the structure of the system, in practice it is still sufficiently severe that we must settle for relatively small values such as $K=8$.

Practicality notwithstanding, we may still inquire as to what would happen in the limit $K \to \infty$. 
In this case, the set of $S$ states converges to a continuum, each one associated with the real-valued time since the node last became susceptible. The aging or decay between these states becomes deterministic, and the linear equations corresponding to finite $K$ become differential equations.

For the random $(q+1)$-regular graph, all nodes and edges are statistically identical.
To describe the limit $K \to \infty$ we use the following quantities and functions, which may vary with time $t$:
\begin{itemize}
\item $S$ and $I$ are the fraction of vertices that are susceptible and infectious respectively.
\item $g(x)$ is the density of vertices that are currently susceptible and have age $x$, i.e., which recovered and became susceptible at time $t-x$.
\item $f(x,y)$ is the density of edges $(i,j)$ where $i$ and $j$ are both susceptible and have ages $x$ and $y$ respectively. We treat $i,j$ and $x,y$ as ordered pairs and count a given edge whose endpoints have ages $x,y$ half towards $f(x,y)$ and half towards $f(y,x)$. Thus $f(x,y)=f(y,x)$ and the total fraction of edges where both endpoints are susceptible is $\iint \dx \,\dy \,f(x,y)$.
\item $h(x)$ is the density of edges where one endpoint is infectious and the other is susceptible with age $x$, where we view these two as an unordered pair. 
\item $\ell$ is the fraction of edges where both endpoints are infectious.
\end{itemize}
We have several identities:
\begin{gather}
S+I = 1 \\
S = \int \dx \,g(x)  \\
\iint \dx \,\dy \,f(x,y) + \int \dx \,h(x) + \ell = 1 \\
g(x) = \int \dy \,f(x,y) + \frac{h(x)}{2} \label{eq:id-g} \\
I = \frac{1}{2} \int \dx \,h(x) + \ell \label{eq:id-I} \, .
\end{gather}
To see Eqs.~\eqref{eq:id-g} and~\eqref{eq:id-I}, think of choosing a random node by first choosing a uniformly random edge and then flipping a coin to choose between its two endpoints: this yields a uniformly random node since our network is regular. In Eq.~\eqref{eq:id-g} we use the fact that $f(x,y) = (f(x,y)+f(y,x))/2$. 

We also define $\sigma(x) = h(x)/g(x)$ as a kind of ``surface-to-volume ratio'' of the set of susceptible vertices, as a function of their age $x$. Specifically, if $i$ is susceptible of age $x$, the probability that a random neighbor of $i$ is infectious is $\sigma(x)/2$, namely the fraction of $g(x)$ contributed by $h(x)/2$ in Eq.~\eqref{eq:id-g}. 

The time derivatives of these functions are as follows. 
\begin{widetext}
\begin{align}
\frac{\partial f(x,y)}{\partial t}
=& -\frac{\partial f(x,y)}{\partial x} - \frac{\partial f(x,y)}{\partial y} 
- \beta q \,\frac{\sigma(x)+\sigma(y)}{2} f(x,y) 
+ \frac{1}{2} \big( \delta(x) h(y) + \delta(y) h(x) \big) 
\label{eq:dfdt} \\ 
\frac{\partial h(x)}{\partial t}
=& -\frac{\partial h(x)}{\partial x} 
- \beta \left( 1 + q \frac{\sigma(x)}{2} \right) h(x) 
+ 2 \beta q \int \dy \,\frac{\sigma(y)}{2} f(x,y) 
+  \big( 2 \delta(x) \ell - h(x) \big) 
\label{eq:dhdt} \\
\frac{\partial \ell}{\partial t}
=& \,\beta\! \int \dx \left( 1 + q \frac{\sigma(x)}{2} \right) h(x) - 2  \ell \, . 
\label{eq:delldt}
\end{align}
\end{widetext}
Here $-\partial f/\partial x$, $-\partial f/\partial y$, and $-\partial h/\partial x$ are flow terms keeping track of the aging of susceptible vertices. The terms proportional to $\beta q \sigma/2$ are flows from $f(x,y)+f(y,x)$ to $h(x)$ and from $h(x)$ to $\ell$: if an endpoint $i$ of an edge $(i,j)$ is susceptible with age $x$, then each of its $q$ neighbors outside that edge is infectious with probability $\sigma(x)/2$ and infects $i$ at rate $\beta$. There is an additional flow from $h(x)$ to $\ell$ of rate $\beta$ since the disease can spread from the infectious endpoint to the susceptible one.
Finally, the terms proportional to $\delta(x)$ are recovery events, where an infectious vertex becomes susceptible with age $0$: the flow from $\ell$ to $h(0)$ has rate $2$, and the flow from $h(x)$ to $f(x,0) + f(0,x)$ has rate $1$.

In Appendix~\ref{app:pde_solve} we show that solving for the steady state of Eqs.~\eqref{eq:dfdt}-\eqref{eq:delldt} is equivalent to finding auxiliary functions $w(x)$ and $\phi(x)$ that obey
\begin{align}
	\frac{\mathrm{d} \phi}{\mathrm{d} x} &= - (\beta +1) \phi(x) + \int_{0}^{\infty}\frac{\beta q \phi(u) \phi(\vert x - u \vert)}{(\gamma(u) + \phi(u))w(u) } \,\du \label{eq:steady_state_phi}\\
	\frac{\mathrm{d} w}{\mathrm{d} x} &= \frac{\beta q \phi(x)}{\gamma(x) + \phi(x)} w(x) 
	\label{eq:steady_state_w} \\
	1 &= \frac{\beta}{\phi(0)} \int_{0}^\infty \left(1 + \frac{q \sigma(x)}{2} \right) \frac{\phi(x)}{w(x)} \,\dx. 
	\label{eq:steady_state_norm}
\end{align}
Unfortunately we are unsure how to solve this set of integro-differential equations except by numerical methods; we suggest this as a direction for future work.
We have solved them numerically for a range of $\beta$ and $q$, although we have some reservation about the approach.
In particular, the numerical solver discretizes to a lattice $x_i = i x_{\mathrm{max}} / N$ for $i=0,\dots,N$ and hence introduces artifacts both due to discretization and truncating the integrals at $x_\mathrm{max}$.
Solving for multiple $x_\mathrm{max}$ and extrapolating to $x_\mathrm{max}=\infty$ appears to work well though it is not obvious how to rigorously quantify the numerical error.

\begin{figure}
\centering
\includegraphics[width=0.85\columnwidth]{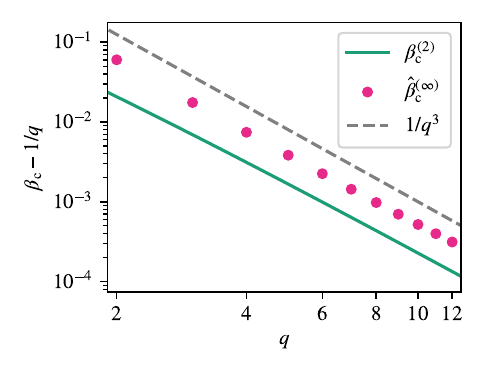}
	\caption{\label{fig:sis_inf} Estimates of the critical spreading rate, $\beta_\mathrm{c}$, for the random $(q+1)$-regular graph.
	The value $\beta_\mathrm{c}^{(2)}$ is computed directly from Eq.~\eqref{eq:beta_c^2}.
	The estimate $\hat{\beta}_\mathrm{c}^{(\infty)}$ is computed by first numerically solving the system in Eqs.~\eqref{eq:steady_state_phi} and \eqref{eq:steady_state_w} and then checking for consistent with Eq.~\eqref{eq:steady_state_norm}. 
	We show the results of a bisection search for the smallest value of $\beta$ that produces a consistent solution.
	}
\end{figure}

To explore Eqs.~\eqref{eq:steady_state_phi}--\eqref{eq:steady_state_norm}, 
we extract predictions for $\beta_\mathrm{c}$ for $q=2,3,\dots,12$.
Results of this exercise are shown in Fig.~\ref{fig:sis_inf}.
Of particular note, recall that we found (analytically) that $\beta_\mathrm{c}^{(2)} = 1/q + 1/4q^3 + \mathcal{O}(1/q^{4})$.
In agreement with this the numerical solutions to the PDE appear consistent with
\begin{equation}
	\beta_\mathrm{c}^{(\infty)} = \frac{1}{q} + \frac{c}{q^3} + \mathcal{O}\Big(\frac{1}{q^4} \Big)
\end{equation}
for some $c>0$. We conjecture that this is the true scaling for the critical value of the SIS model on random $(q+1)$-regular graphs: that is, in a large-degree approximation where we expand $\beta_c$ in powers of $1/q$, we conjecture that $c/q^3$ is the first nonzero correction.

Finally, while these equations describe the large-$K$ limit of our approach, we believe that that they are still not exact, even on random regular graphs. 
In particular, we have only incorporated one kind of memory---the time since a node last recovered and became susceptible.
If we further augmented the state space to track more past infections and recoveries at each node, we should approach the continuum limit of the dynamical cavity method.

\section{Discussion}

By augmenting the state space to keep track of how long a node has been susceptible, and applying the pair approximation to these augmented states, we have derived a set of fixed-point equations to accurately approximate the behavior of the SIS model.
On any finite network of interest, these equations can be iterated to convergence in order to make predictions without simulating the process, and thus approximate both the epidemic threshold and the endemic fraction of individuals.  
Additionally, on ensembles of networks such as random regular graphs or random graphs with a given degree distribution, these equations can be solved to approximate the SIS model on the corresponding ensemble.

Increasing the number $K$ of susceptible states rapidly increases the accuracy of our approximation. 
As evidence for this, we compared our results to numerical simulation on random graphs and two real-world networks.
In all cases we found good  agreement between predictions and simulations for $K=8$.

Our approach focuses on temporal correlations, namely between the times that neighboring nodes have been susceptible and the probability their neighbors are infected. The computational cost of this method increases polynomially with $K$. 
In contrast, modeling longer-range spatial correlations using $k$-tuples of nodes   has complexity exponential in $k$. Additionally, we see no obvious and unique decomposition for arbitrary networks into sets of $k$-tuples, while the decomposition for the pair approximation simply corresponds to the set of edges.

While our approximations are very accurate in practice, we hypothesize that they are not exact even in the limit $K \to \infty$. 
In order to improve them further one could introduce additional states to keep track of more previous infection and recovery transitions, and thus capture more temporal correlations between a node's recent history and those of its neighbors. However, as in the dynamical cavity method the size of the state space would grow exponentially with the number of transitions.
We leave this for future work.

Finally, we have focused on the SIS model but it seems hopeful that the the technique we advanced will generalize to other settings, such as more sophisticated epidemic models, dynamic spin models, or lattice gases.
This, again, we leave to future work.\hfill \break

\paragraph*{\bf Acknowledgements.}
C.M. acknowledges support from National Science Foundation Grant BIGDATA-1838251 and the Robert Wood Johnson Foundation, and the hospitality of the Sea Cloud II.
G.T.C. thanks Guillaume St-Onge for useful conversations.
Both thank an anonymous reviewer for tips on how to solve systems of PDEs.
\hfill \break

\paragraph*{\bf Code availability.} Computer code implementing our methods is available at
\href{https://github.com/gcant/SISK}{github.com/gcant/SISK}.

\bibliographystyle{numeric}

\begin{thebibliography}{10}
\expandafter\ifx\csname url\endcsname\relax
  \def\url#1{\texttt{#1}}\fi
\expandafter\ifx\csname urlprefix\endcsname\relax\def\urlprefix{URL }\fi

\bibitem{kiss_mathematics_2017}
I.~Z. Kiss, J.~C. Miller, and P.~L. Simon, \textit{Mathematics of {Epidemics}
  on {Networks}: {From} {Exact} to {Approximate} {Models}}, volume~46 of
  \textit{Interdisciplinary {Applied} {Mathematics}}. Springer International
  Publishing, Cham (2017),
  \urlprefix\url{http://link.springer.com/10.1007/978-3-319-50806-1}.

\bibitem{pastor-satorras_epidemic_2015}
R.~Pastor-Satorras, C.~Castellano, P.~Van~Mieghem, and A.~Vespignani, Epidemic
  processes in complex networks. \textit{Reviews of Modern Physics}
  \textbf{87}(3), 925--979 (2015),
  \urlprefix\url{https://link.aps.org/doi/10.1103/RevModPhys.87.925}.

\bibitem{keeling_networks_2005}
M.~J. Keeling and K.~T. Eames, Networks and epidemic models. \textit{Journal of
  The Royal Society Interface} \textbf{2}(4), 295--307 (2005),
  \urlprefix\url{https://royalsocietypublishing.org/doi/10.1098/rsif.2005.0051}.

\bibitem{grassberger_critical_1983}
P.~Grassberger, On the critical behavior of the general epidemic process and
  dynamical percolation. \textit{Mathematical Biosciences} \textbf{63}(2),
  157--172 (1983),
  \urlprefix\url{https://linkinghub.elsevier.com/retrieve/pii/0025556482900360}.

\bibitem{newman_spread_2002}
M.~E.~J. Newman, Spread of epidemic disease on networks. \textit{Physical
  Review E} \textbf{66}(1), 016128 (2002),
  \urlprefix\url{https://link.aps.org/doi/10.1103/PhysRevE.66.016128}.

\bibitem{moore_epidemics_2000}
C.~Moore and M.~E.~J. Newman, Epidemics and percolation in small-world
  networks. \textit{Physical Review E} \textbf{61}(5), 5678--5682 (2000),
  \urlprefix\url{https://link.aps.org/doi/10.1103/PhysRevE.61.5678}.

\bibitem{Diekmann_1998}
O.~Diekmann, {{M. C. M. De Jong}}, and J.~A.~J. Metz, A deterministic epidemic
  model taking account of repeated contacts between the same individuals.
  \textit{Journal of Applied Probability} \textbf{35}(2), 448--462 (1998),
  \urlprefix\url{http://www.jstor.org/stable/3215698}.

\bibitem{pastor-satorras_epidemic_2001}
R.~Pastor-Satorras and A.~Vespignani, Epidemic spreading in scale-free
  networks. \textit{Physical Review Letters} \textbf{86}, 3200--3203 (2001),
  \urlprefix\url{https://link.aps.org/doi/10.1103/PhysRevLett.86.3200}.

\bibitem{harris_contact_1974}
T.~E. Harris, Contact interactions on a lattice. \textit{The Annals of
  Probability} \textbf{2}(6), 969 -- 988 (1974),
  \urlprefix\url{https://doi.org/10.1214/aop/1176996493}.

\bibitem{weiss_asymptotic_1971}
G.~H. Weiss and M.~Dishon, On the asymptotic behavior of the stochastic and
  deterministic models of an epidemic. \textit{Mathematical Biosciences}
  \textbf{11}(3-4), 261--265 (1971),
  \urlprefix\url{https://linkinghub.elsevier.com/retrieve/pii/0025556471900873}.

\bibitem{1238052}
Y.~Wang, D.~Chakrabarti, C.~Wang, and C.~Faloutsos, Epidemic spreading in real
  networks: an eigenvalue viewpoint. In \textit{22nd International Symposium on
  Reliable Distributed Systems, 2003. Proceedings.}, pp. 25--34 (2003),
  \urlprefix\url{https://doi.org/10.1109/RELDIS.2003.1238052}.

\bibitem{chakrabarti_epidemic_2008}
D.~Chakrabarti, Y.~Wang, C.~Wang, J.~Leskovec, and C.~Faloutsos, Epidemic
  thresholds in real networks. \textit{ACM Transactions on Information and
  System Security} \textbf{10}(4), 1--26 (2008),
  \urlprefix\url{https://dl.acm.org/doi/10.1145/1284680.1284681}.

\bibitem{4549746}
P.~Van~Mieghem, J.~Omic, and R.~Kooij, Virus spread in networks.
  \textit{IEEE/ACM Transactions on Networking} \textbf{17}(1), 1--14 (2009),
  \urlprefix\url{https://dl.acm.org/doi/10.1109/TNET.2008.925623}.

\bibitem{van2011n}
P.~Van~Mieghem, The {N}-intertwined {SIS} epidemic network model.
  \textit{Computing} \textbf{93}(2), 147--169 (2011),
  \urlprefix\url{https://doi.org/10.1007/s00607-011-0155-y}.

\bibitem{keeling_1999}
M.~J. Keeling, The effects of local spatial structure on epidemiological
  invasions. \textit{Proceedings of the Royal Society of London. Series B:
  Biological Sciences} \textbf{266}(1421), 859--867 (1999),
  \urlprefix\url{https://royalsocietypublishing.org/doi/abs/10.1098/rspb.1999.0716}.

\bibitem{mata_pair_2013}
A.~S. Mata and S.~C. Ferreira, Pair quenched mean-field theory for the
  susceptible-infected-susceptible model on complex networks.
  \textit{Europhysics Letters} \textbf{103}(4), 48003 (2013),
  \urlprefix\url{https://iopscience.iop.org/article/10.1209/0295-5075/103/48003}.

\bibitem{shrestha2015message}
M.~Shrestha, S.~V. Scarpino, and C.~Moore, Message-passing approach for
  recurrent-state epidemic models on networks. \textit{Physical Review E}
  \textbf{92}(2), 022821 (2015),
  \urlprefix\url{https://doi.org/10.1103/PhysRevE.92.022821}.

\bibitem{cator_second-order_2012}
E.~Cator and P.~Van~Mieghem, Second-order mean-field
  susceptible-infected-susceptible epidemic threshold. \textit{Physical Review
  E} \textbf{85}(5), 056111 (2012),
  \urlprefix\url{https://link.aps.org/doi/10.1103/PhysRevE.85.056111}.

\bibitem{gleeson_2011}
J.~P. Gleeson, High-accuracy approximation of binary-state dynamics on
  networks. \textit{Physical Review Letters} \textbf{107}, 068701 (2011),
  \urlprefix\url{https://link.aps.org/doi/10.1103/PhysRevLett.107.068701}.

\bibitem{karrer2010message}
B.~Karrer and M.~E.~J. Newman, Message passing approach for general epidemic
  models. \textit{Physical Review E} \textbf{82}(1), 016101 (2010),
  \urlprefix\url{https://doi.org/10.1103/PhysRevE.82.016101}.

\bibitem{karrer_percolation_2014}
B.~Karrer, M.~E.~J. Newman, and L.~Zdeborov\'a, Percolation on sparse networks.
  \textit{Phys. Rev. Lett.} \textbf{113}, 208702 (2014),
  \urlprefix\url{https://link.aps.org/doi/10.1103/PhysRevLett.113.208702}.

\bibitem{ferreira_critical_2013}
R.~S. Ferreira and S.~C. Ferreira, Critical behavior of the contact process on
  small-world networks. \textit{The European Physical Journal B}
  \textbf{86}(11), 462 (2013),
  \urlprefix\url{http://link.springer.com/10.1140/epjb/e2013-40534-0}.

\bibitem{gleeson_2013}
J.~P. Gleeson, Binary-state dynamics on complex networks: Pair approximation
  and beyond. \textit{Physical Review X} \textbf{3}(2), 021004 (2013),
  \urlprefix\url{https://link.aps.org/doi/10.1103/PhysRevX.3.021004}.

\bibitem{zhang_inference_2012}
P.~Zhang, Inference of kinetic {I}sing model on sparse graphs. \textit{Journal
  of Statistical Physics} \textbf{148}(3), 502--512 (2012),
  \urlprefix\url{http://link.springer.com/10.1007/s10955-012-0547-1}.

\bibitem{PhysRevE.91.012811}
A.~Y. Lokhov, M.~M\'ezard, and L.~Zdeborov\'a, Dynamic message-passing
  equations for models with unidirectional dynamics. \textit{Physical Review E}
  \textbf{91}, 012811 (2015),
  \urlprefix\url{https://link.aps.org/doi/10.1103/PhysRevE.91.012811}.

\bibitem{Dominguez_2020}
E.~Domínguez, D.~Machado, and R.~Mulet, The cavity master equation: average
  and fixed point of the ferromagnetic model in random graphs. \textit{Journal
  of Statistical Mechanics: Theory and Experiment} \textbf{2020}(7), 073304
  (2020), \urlprefix\url{https://dx.doi.org/10.1088/1742-5468/ab9eb6}.

\bibitem{Aurell_2023}
E.~Aurell, D.~Machado~Perez, and R.~Mulet, A closure for the master equation
  starting from the dynamic cavity method. \textit{Journal of Physics A:
  Mathematical and Theoretical} \textbf{56}(17), 17LT02 (2023),
  \urlprefix\url{https://dx.doi.org/10.1088/1751-8121/acc8a4}.

\bibitem{PhysRevX.13.031021}
F.~Behrens, B.~Hudcov\'a, and L.~Zdeborov\'a, Backtracking dynamical cavity
  method. \textit{Phys. Rev. X} \textbf{13}, 031021 (2023),
  \urlprefix\url{https://link.aps.org/doi/10.1103/PhysRevX.13.031021}.

\bibitem{PhysRevE.105.024308}
E.~Ortega, D.~Machado, and A.~Lage-Castellanos, Dynamics of epidemics from
  cavity master equations: Susceptible-infectious-susceptible models.
  \textit{Physical Review E} \textbf{105}, 024308 (2022),
  \urlprefix\url{https://link.aps.org/doi/10.1103/PhysRevE.105.024308}.

\bibitem{Norris_1997}
J.~R. Norris, \textit{Markov Chains}. Cambridge Series in Statistical and
  Probabilistic Mathematics, Cambridge University Press (1997).

\bibitem{morris_hiv_2011}
M.~Morris and R.~Rothenberg, {HIV} {Transmission} {Network} {Metastudy}
  {Project}: {An} {Archive} of {Data} {From} {Eight} {Network} {Studies},
  1988--2001: {Version} 1 (2011),
  \urlprefix\url{https://www.icpsr.umich.edu/web/NAHDAP/studies/22140/versions/V1}.

\bibitem{subelj_robust_2011}
L.~Šubelj and M.~Bajec, Robust network community detection using balanced
  propagation. \textit{The European Physical Journal B} \textbf{81}(3),
  353--362 (2011),
  \urlprefix\url{http://link.springer.com/10.1140/epjb/e2011-10979-2}.

\bibitem{spectral_redemption_2013}
F.~Krzakala, C.~Moore, E.~Mossel, J.~Neeman, A.~Sly, L.~Zdeborová, and
  P.~Zhang, Spectral redemption in clustering sparse networks.
  \textit{Proceedings of the National Academy of Sciences} \textbf{110}(52),
  20935--20940 (2013),
  \urlprefix\url{https://www.pnas.org/doi/abs/10.1073/pnas.1312486110}.

\bibitem{MacDonald78}
N.~MacDonald, \textit{Time Lags in Biological Models}. Lecture Notes in Biomathematics, Springer Berlin (1978),
  \urlprefix\url{https://doi.org/10.1007/978-3-642-93107-9}.


\end{thebibliography}

\appendix

\section{Linear stability for pair approximations}
\label{app:evalue}
First we derive a result, reminiscent of the Ihara-Bass formula~\cite{spectral_redemption_2013}, for the eigenvalues of a modified non-backtracking operator.
Specifically, let $z_{ij}$ be an eigenvector of the operator
\begin{equation}
z_{ij} = w B_{ij}(\boldsymbol{z}) + v B_{ji}(\boldsymbol{z})
\label{eq:bnb_eigenvector}
\end{equation}
and define
\begin{equation}
\mu_j = \sum_{k}A_{jk} z_{jk} \, .
\end{equation}
Then
\begin{equation}
z_{ji} = \frac{w}{1+v}(\mu_i - z_{ij}) + \frac{v}{1+v} \,\mu_j
\label{eq:ihara_1}
\end{equation}
and
\begin{equation}
\sum_j A_{ij} z_{ji} = \frac{w}{1+v}(d_i - 1)\mu_i + \frac{v}{1+v} \sum_j A_{ij} \mu_j
\label{eq:ihara_2}
\end{equation}
where $d_i = \sum_j A_{ij}$ is the degree of node~$i$.

Now consider the quantity
\begin{equation}
\begin{aligned}
\sum_{j} A_{ij} \mu_j &= \sum_{j} A_{ij} \sum_{k} A_{jk} z_{jk} \\
&= \sum_j A_{ij} \big( z_{ji} + B_{ij}(\boldsymbol{z}) \big) \\
&= \sum_j A_{ij} \Big( z_{ji} + \frac{1}{w} z_{ij} - \frac{v }{ w} B_{ji}(\boldsymbol{z}) \Big) \\
&= \sum_j A_{ij} \Big( z_{ji} + \frac{1 +v}{w} z_{ij} - \frac{v}{w}  \mu_i \Big)
\, .
\end{aligned}
\end{equation}
Using Eqs.~\eqref{eq:ihara_1} and \eqref{eq:ihara_2} we arrive at
\begin{equation}
\left( \Big( \frac{w}{1+v} -\frac{v}{w}\Big)(D-I) - \frac{1}{1+v} A + \frac{1}{w} I   \right) \boldsymbol{\mu}
= \boldsymbol{0}
\end{equation}
where $D$ is the diagonal matrix of node degrees, $D_{ii} = d_i$.

For the pair approximation to the SIS model we compare Eq.~\eqref{eq:bnb_eigenvector} to Eq.~\eqref{eq:linear_Psi} and
set 
\[
w=\frac{\beta(\beta+2)}{2(\beta+1)} 
\quad \text{and} \quad v=\frac{\beta^2}{2(\beta+1)} \, ,
\]
and let $D = L + A$ for the Laplacian $L$ to get
\begin{equation}
\left( I + \beta \Big(I + \frac{\beta}{2}L - A \Big) \right) \boldsymbol{\mu} = \boldsymbol{0} 
\, .
\end{equation}
Then the prediction for the critical value of $\beta$ is the value that solves
\begin{equation}
\beta = \frac{1}{\lambda\big(A - {\beta} L/2 - I \big) }
\, .
\end{equation}

\section{Steady state of the PDEs for the $K \to \infty$ limit}
\label{app:pde_solve}

To find the steady state solution (assuming it exists), we set all time derivatives to zero in Eqs.~\eqref{eq:dfdt}-\eqref{eq:delldt}.
We also define the following three auxiliary functions:
\begin{align}
	w(x) &= \exp \left( \frac{\beta q}{2} \int_{0}^x \sigma(y) \,\dy \right) \\
	\phi(x) &= h(x) w(x) \\
	\gamma(x) &= \int_{0}^\infty \frac{\phi(\vert x - y \vert)}{w(y)} \,\dy.
\end{align}

To make progress, let us first consider the value of $f(x,y)$ along lines $(x,y) = (x_0+s, y_0+s)$.
Setting $\partial f / \partial t = 0$ in Eq.~\eqref{eq:dfdt} gives 
\begin{equation}
	\frac{\mathrm{d} f}{\mathrm{d} s} = -\beta q  \frac{\sigma(x_0 + s) + \sigma(y_0 + s)}{2} f(x_0+s, y_0+s)
\end{equation}
which yields
\begin{equation}
	\frac{f(x_0 +s, y_0+s)}{f(x_0, y_0)} = \e^{-\frac{\beta q}{2} \int_0^s (\sigma(x_0+u) + \sigma(y_0+u)) \,\du} \, .
\end{equation}
For arbitrary $x > y$, set $s=y$, $x_0=x-y$ and $y_0=0$ to find 
\begin{align}
	f(x, y) = f(x-y, 0) \,\frac{w(x-y)}{w(x)w(y)}
    \, .
\end{align}
Then, noting the boundary condition  $f(x,0) = h(x)/2$ and the symmetry in $x$ and $y$, we get the steady-state solution
\begin{equation}
	f(x, y) = \frac{\phi(\vert x - y \vert)}{2 w(x) w(y)}
    \, . 
    \label{eq:steady_state_f}
\end{equation}
Hence if we can find $\phi(x)$ and $w(x)$ then we can immediately compute $f$, $h$ and $\ell$.

From this equation for $f(x,y)$ we see that
\begin{equation}
\sigma(x) = \frac{2 h(x)}{2 \int_0^\infty f(x,y) \,\dy + h(x)}=\frac{2 \phi(x)}{\gamma(x) + \phi(x)}
\end{equation}
and since $w' = \beta q \sigma w / 2$ we have that
\begin{align}
	\frac{\mathrm{d}\phi}{\mathrm{d}x} &= \frac{\mathrm{d}h}{\mathrm{d}x} \,w(x) + \frac{\mathrm{d}w}{\mathrm{d}x} \,h(x) \nonumber \\
	&= w(x) \left(\frac{\mathrm{d}h}{\mathrm{d}x} + \frac{\beta q \sigma(x)}{2} \,h(x)  \right) .
\end{align}
Inserting the identity for $\mathrm{d}h / \mathrm{d}x$ from Eq.~\eqref{eq:dhdt} (again, setting $\partial h/ \partial t=0$) we get
\begin{align}
	\frac{\mathrm{d}\phi}{\mathrm{d}x} &= - (\beta +1) \phi(x) + \frac{\beta q}{2} \int_{0}^{\infty}\frac{\sigma(u) \,\phi(\vert x - u \vert)}{w(u)} \,\du \\
	\frac{\mathrm{d}w}{\mathrm{d}x} &= \frac{\beta q}{2} \,\sigma(x) w(x)
	\label{eq:phi-w-diff-eq}
\end{align}
with the boundary condition that $\phi(0) = 2 \ell$.

Hence, if we can find $\phi(x)$ and $w(x)$ that solve these equations then we have putative expressions for the steady state $f$ and $h$.
However, we should also check that the equation for $\ell$ is satisfied, namely, 
\begin{equation}
	\frac{\beta}{\phi(0)} \int_{0}^\infty \left(1 + \frac{q \sigma(x)}{2} \right) \frac{\phi(x)}{w(x)} \,\dx = 1 \, .
	\label{eq:self_consistent_normalization}
\end{equation}
If this holds, we have a complete set of equations for the steady-state.

To solve these equations we
iteratively update estimates for $w(x)$ and $\widetilde{\phi}(x) = \phi(x) / \phi(0)$.
(Note: $\widetilde{\phi}(x)$ obeys the same differential equation as $\phi(x)$ but conveniently has $\widetilde{\phi}(0)=1$, so that we do not need to track $\ell$ inside the iteration.)
The algorithm is as follows.
\begin{center}
\fbox{%
  \begin{minipage}{0.90\columnwidth}
    \textbf{initialization:} 
    \begin{equation*}
    \widetilde{\phi}^{(0)}(x) = \e^{-(\beta +1)x} \quad\text{and}\quad  w^{(0)}(x) = 1
    \end{equation*}
    \textbf{update:}
    \begin{align*}
    \widetilde{\gamma}^{(t)}(x) &= \int_{0}^{\infty} \bigg( \frac{\widetilde{\phi}^{(t)}(\vert x - y \vert)}{w^{(t)}(y)}\bigg) \dy \\
    \sigma^{(t)}(x) &= \frac{2 \widetilde{\phi}^{(t)}(x)}{\widetilde{\gamma}^{(t)}(x) + \widetilde{\phi}^{(t)}(x)} \\
    c^{(t)}(x) &= \frac{\beta q}{2} \int_{0}^{\infty} \bigg( \frac{ \widetilde{\phi}^{(t)}(\vert x - y \vert)\, \sigma^{(t)}(y)}{w^{(t)}(y)} \bigg)\dy \\
    w^{(t+1)}(x) &= \exp\Big( {\frac{\beta q}{2} \int_{0}^{x} \sigma^{(t)}(y) \, \dy }\Big) \\
    \widetilde{\phi}^{(t+1)}(x) &= \e^{-(\beta+1)x} \Big( 1 + \int_{0}^{x} \e^{(\beta+1)y}\, c^{(t)}(y)\, \dy \Big)
    \end{align*}
  \end{minipage}%
}
\end{center}

To compute the required quantities numerically we discretize to a uniform grid $x_i = i x_{\mathrm{max}} / N$ for $i = 0, \dots, N$.
The convolutions for $\gamma^{(t)}$ and $c^{(t)}$ are approximated using a fast Fourier transform and the integrals for $w^{(t+1)}$ and $\widetilde{\phi}^{(t+1)}$ are approximated using Simpon's rule.
After the convergence of $\widetilde{\phi}(x)$ and $w(x)$, we check to see if Eq.~\eqref{eq:self_consistent_normalization} is obeyed. 
If so, we have found a non-trivial steady state.
If~\eqref{eq:self_consistent_normalization} is violated we assume that a non-trivial steady state does not exist, i.e., the process is sub-critical.

Each value of $x_\mathrm{max}$ makes slightly different predictions.
To find the critical $\beta$, we re-run calculations for a range of $x_\mathrm{max}$ and extrapolate Eq.~\eqref{eq:self_consistent_normalization} by fitting
\begin{equation}
	\beta \int_{0}^{x_\mathrm{max}} \left( 1 + \frac{q \sigma(x)}{2} \right) \frac{\widetilde{\phi}(x)}{w(x)} \,\dx = \alpha_0 + \frac{\alpha_1}{x_\mathrm{max}}
\end{equation}
and if $\alpha_0 < 1$ we assume the process is still sub-critical when $x_{\mathrm{max}} \to \infty$. This results in the estimates of $\beta_c^{(\infty)}$ shown in Fig.~\ref{fig:sis_inf}.

\end{document}